\documentclass[showpacs,aps,prb,reprint,superscriptaddress,nofootinbib,
longbibliography]{revtex4-2}
\usepackage[colorlinks=true, pdfstartview=FitV,
bookmarks=true, bookmarksnumbered=true, breaklinks]{hyperref}
\usepackage[dvipdfmx]{graphicx}
\usepackage{amsmath,amssymb,bm,color,longtable,mathrsfs,slashed,comment,tikz}
\definecolor{darkpastelgreen}{rgb}{0.01, 0.75, 0.24}
\definecolor{electricindigo}{rgb}{0.44, 0.0, 1.0}
\definecolor{palatinateblue}{rgb}{0.15, 0.23, 0.89}
\definecolor{carminered}{rgb}{1.0, 0.0, 0.22}
\hypersetup{linkcolor=palatinateblue, citecolor=electricindigo, urlcolor=carminered}

\newcommand{\shorten}[1]{\if 0 #1 \fi}

\newcommand{\suppA}
{Appendix~\ref{sec:mixing_term} {}}
\newcommand{\suppB}
{Appendix~\ref{sec:gap-eqs} {}}

\newcommand{\sect}[1]{\section{#1}}

\newcommand{ \arcsinh}{{\rm arcsinh}}
\newcommand{\R}{\Phi}
\renewcommand{\=}{&=&}

\newcommand{\cout}[1]{ \if 0 {#1} \fi }

\newcommand{\nn}{\nonumber}
\renewcommand{\=}{&=&}
\newcommand{\nnb}{\nonumber \\}
\newcommand{\pd}{\partial}

\newcommand{\sla}{ \slashed }

\newcommand{\ep}{ \epsilon }

\newcommand{\gam}{ \gamma }

\newcommand{\Lag}{ {\mathcal{L}} }

\newcommand{\order}{{\mathcal O}}
\newcommand{\prj}{ {\mathcal P} }
\newcommand{\Q}{ {\mathcal Q} }

\newcommand{\para}{ \parallel}

\newcommand{\sgn}{ {\rm sgn} }

\newcommand{\vac}{ {\rm vac} }

\newcommand{\com}[1]{[{\color[rgb]{0,0,1}{#1}}]}

\newcommand{\cgd}[1]{{\color[rgb]{1,0,0}{#1}}}

\newcommand{\gr}[1]{{\color[rgb]{0.3,0.6,0.3}{#1}}}

\newcommand{\pp}[1]{{\color[rgb]{0.7,0.3,0.9}{#1}}}

\definecolor{lime}{HTML}{A6CE39}
\DeclareRobustCommand{\orcidicon}{
	\hspace{-3mm}
	\begin{tikzpicture}
	\draw[lime, fill=lime] (0,0) 
	circle [radius=0.16] 
	node[white] {{\fontfamily{qag}\selectfont \tiny ID}};
	\draw[white, fill=white] (-0.0625,0.095) 
	circle [radius=0.007];
	\end{tikzpicture}
	\hspace{-3mm}
}

\foreach \x in {A, ..., Z}{\expandafter\xdef\csname orcid\x\endcsname{\noexpand\href{https://orcid.org/\csname orcidauthor\x\endcsname}
			{\noexpand\orcidicon}}
}

\begin{document}

\title{ 
Dirac Kondo effect under magnetic catalysis
}

\author{Koichi~Hattori\orcidA{}}
\email[]{koichi.hattori@zju.edu.cn}
\affiliation{Zhejiang Institute of Modern Physics, Department of Physics, 
Zhejiang University, Hangzhou, 310027, China} 
\affiliation{Research Center for Nuclear Physics (RCNP), Osaka University, Osaka 567-0047, Japan}

\author{Daiki~Suenaga\orcidB{}}
\email[]{daiki.suenaga@riken.jp}
\affiliation{Strangeness Nuclear Physics Laboratory, RIKEN Nishina Center, Wako 351-0198, Japan}
\affiliation{Research Center for Nuclear Physics (RCNP), Osaka University, Osaka 567-0047, Japan}

\author{Kei~Suzuki\orcidC{}}
\email[]{{k.suzuki.2010@th.phys.titech.ac.jp}}
\affiliation{Advanced Science Research Center, Japan Atomic Energy Agency (JAEA), Tokai 319-1195, Japan}

\author{Shigehiro~Yasui\orcidD{}}
\email[]{yasuis@keio.jp}
\affiliation{International Institute for Sustainability with Knotted Chiral Meta Matter (WPI-SKCM$^{2}$), Hiroshima University, Higashi-Hiroshima, Hiroshima 739-8526, Japan}
\affiliation{Research and Education Center for Natural Sciences, Keio University, Hiyoshi 4-1-1, Yokohama, Kanagawa 223-8521, Japan}

\date{\today}

\begin{abstract}
We develop a mean-field theory of a novel Kondo effect emerging in systems without a Fermi surface, which instead emerges under strong magnetic fields.
We determine the magnitude of the Kondo condensate 
which is a particle pairing composed of 
conducting Dirac fermions and localized impurities. 
We focus on the competition between the Kondo effect and the energy gap formation that stems from the pairing among the Dirac fermions leading to the dynamical chiral symmetry breaking.  
We find that this competition induces a quantum critical point. 
We also investigate finite-temperature effects. 
This system at vanishing fermion density can be studied with Monte Carlo lattice simulations which do not suffer from the sign problem. 
\end{abstract}

\maketitle

\sect{Introduction} \label{Sec:1}
Quantum systems under strong magnetic fields have attracted much attention over the last century in various physical systems (see, e.g., Refs.~\cite{Grasso:2000wj, Dittrich:2000zu, yoshioka2002quantum, Dunne:2004nc, Harding:2006qn, Mourou:2006zz}
for reviews). 
In the last decade, even more, common interests have been developed with the aid of advanced engineering of ultrarelativistic heavy-ion collisions \cite{Kharzeev:2013jha, Kharzeev:2013ffa, Kharzeev:2015znc, Hattori:2016emy, Koch:2016pzl} 
and of Dirac/Weyl semimetals \cite{Burkov:2015hba, Kharzeev:2015kna, Miransky:2015ava, Armitage:2017cjs} 
as well as the tremendous progress in the numerical lattice simulations \cite{DElia:2012ems, Endrodi:2015oba} (see also references therein). 
These situations motivate us to study phases of matter under strong magnetic fields. 
As we will discuss in detail, 
it has been pointed out that 
strong magnetic fields {\it catalyze} particle pairing phenomena, inducing nonpertubative modification of the ground state of the system when the Landau quantization becomes sizable.

In this paper, we investigate an interacting system composed of relativistic light fermions and nonrelativistic heavy impurities under a strong magnetic field at vanishing fermion density.  
Such relativistic fermions, governed by the Dirac equation, 
appear not only as elementary particles in high-energy physics 
but also in Dirac semimetals in condensed matter physics. 
Light and heavy fermions correspond to conducting and localized states, respectively. 
The conventional Kondo effect is absent at vanishing fermion density, whereas our system exhibits the Kondo effect and the chiral symmetry breaking induced by strong external magnetic fields. 
We call the Kondo effect appearing in Dirac fermion systems the {\it Dirac Kondo effect}. 
These phenomena are induced by two pairing patterns: the Kondo condensate, pairing between the light fermion and the heavy impurity, and the chiral condensate, pairing between the light fermion and its antiparticle. 
In fact, each condensate is induced even with infinitesimal attractive interactions between the pair. 
That is, the growth of each condensate is an inevitable fate 
of the system when a magnetic field increases and temperature decreases. 
The common concept behind this statement is the {\it magnetic catalysis} 
as first pointed out in the formation of the chiral condensate~\cite{Gusynin:1994re,Gusynin:1994va,Gusynin:1994xp,Gusynin:1995nb} and refined in the formation of the Kondo condensate~\cite{Ozaki:2015sya}.\footnote{
While the magnetic catalysis is conventionally meant for the chiral condensate, the key concept, 
the {\it dimensional reduction} in the phase space stemming from the Landau degeneracy, 
is found to be more general and 
can lead to the realization of various pairing patterns 
in analogy with the Fermi surface effect \cite{Polchinski:1992ed, Shankar:1993pf}. 
}

This system provides an interesting question arising from the competition between the chiral and Kondo condensates, which can drastically change the fate of the system.  
We will find that one condensate interrupts the growth of the other destructively, and there occurs a transition from the vacuum completely dominated by the chiral condensate 
to that accompanied by the Kondo condensate. 
At zero temperature, these two phases are 
divided by a quantum critical point 
at a certain critical magnetic-field strength 
where the Kondo condensate starts to grow. 
The chiral condensate no longer grows above 
the critical point and saturates at a constant value.  
\shorten{
We also investigate finite-temperature effects. 
This system at vanishing fermion density can be studied with the Monte Carlo lattice simulations which do not suffer from the sign problem. 
}

Our findings have an impact on the heart of
many-body quantum physics. 
It is a general and central issue to determine the ground state of a system with condensates. 
The ground states can exhibit nonperturbative quantum phenomena such as superconductivity 
\cite{Bardeen:1957mv, Polchinski:1992ed, Shankar:1993pf} 
and the Kondo effect \cite{Kondo:1964,PhysicsPhysiqueFizika.2.5, anderson1970poor, Wilson:1974mb, Hewson,Yosida,Yamada,coleman_2015}. 
Relativistic counterparts are 
the spontaneous chiral symmetry breaking \cite{Nambu:1961tp, Nambu:1961fr, Klevansky:1992qe, Hatsuda:1994pi}, 
the color superconductivity \cite{IWASAKI1995163, Alford:1997zt, Rapp:1997zu, Alford:2007xm, Fukushima:2010bq}, 
and the newly proposed QCD Kondo effect \cite{Yasui:2013xr,Hattori:2015hka} (see also Refs.~\cite{Kanazawa:2016ihl,Kimura:2016zyv,Yasui:2016svc,Yasui:2016yet,Yasui:2017izi,Suzuki:2017gde,Yasui:2017bey,Kimura:2018vxj,Fariello:2019ovo,Hattori:2019zig,Suenaga:2019car,Suenaga:2019jqu,Kanazawa:2020xje,Araki:2020fox,Araki:2020rok,Kimura:2020uhq,Suenaga:2020oeu,Ishikawa:2021bey,Suenaga:2021wio} for the developments in the QCD Kondo effect). 
Relativistic systems can be also realized with 
graphene and Dirac/Weyl semimetals 
that provide useful platforms in the study of the magnetic catalysis \cite{Semenoff:1999xv, Semenoff:1998bk, khveshchenko2001magnetic, Gorbar:2002iw, Gusynin:2006gn, herbut2007theory, Herbut:2008ui, gusynin2009edge, roy2011inhomogeneous, roy2014chiral, 
Roy:2014mia, PhysRevB.95.205108, PhysRevB.96.235104, 
DeTar:2016vhr, PhysRevB.89.245404, PhysRevB.95.165442, PhysRevX.6.011029, tada2020quantum} 
and the Kondo effect~\cite{Principi:2015,Yanagisawa:2015conf,Yanagisawa:2015,Mitchell:2015,Sun:2015,Feng:2016,Kanazawa:2016ihl,Lai:2018,Ok:2017,PhysRevB.97.045148,PhysRevB.98.075110,Dzsaber:2018,PhysRevB.99.115109,KIM2019236,Grefe:2019,Grefe:2020,Pedrosa:2021} (see Refs.~\cite{Miransky:2015ava, Fritz_2013} for reviews).\footnote{
References~\cite{KIM2019236, Grefe:2020} studied 
the Zeeman effect on the Kondo effect at {\it finite} density, while we discuss the magnetic catalysis for the Kondo effect at {\it vanishing} density. 
}

Despite the many foregoing studies, it should be stressed  
that the competition of the Kondo condensate 
with other condensates is yet elusive; 
The interplay appears both in destructive and constructive manners.  
There is a destructive competition between the Kondo effect and superconductivity \cite{
soda1967s, fowler1967conditions, *fowler1970conditions, 
zittartz1970theory, *muller1970theory, *muller1971kondo, 
sakurai1970comments, 
matsuura1976depression, *matsuura1977effects, *matsuura1977theory, 
jarrell1990gap, satori1992numerical, *shiba1993numerical, *sakai1993numerical, salkola1997spectral, bauer2013microscopic, kirvsanskas2015yu, Kanazawa:2016ihl, lee2017scaling, moca2021kondo} (see also, e.g., Refs.~\cite{maple1972re, yazdani1997probing, franke2011competition, hatter2015magnetic, balatsky2006impurity} for experiments). 
Two of the present authors have also discussed the competition between the chiral symmetry breaking 
and the QCD Kondo effect at finite baryon density \cite{Suzuki:2017gde} (see also Refs.~\cite{Kanazawa:2020xje, Ishikawa:2021bey}).
On the other hand, constructive interplay can induce  
``heavy fermion'' superconductivity \cite{Hewson, balatsky2006impurity, coleman2006heavy, pfleiderer2009superconducting, stewart2017unconventional}. 
It is an interesting question how the Kondo condensate could affect the critical point from the interplay between the chiral condensate and the color superconductivity \cite{Hatsuda:2006ps, Yamamoto:2007ah, Fukushima:2010bq}.

This paper is organized as follows.
In Sec.~\ref{Sec:2}, we introduce a model Lagrangian and construct an effective potential within the mean-field approximation. 
Based on this, we discuss the phase structure with respect to a magnetic-field strength at zero temperature in Sec.~\ref{Sec:3} and finite-temperature effects at Sec.~\ref{Sec:3-2}. 
In Sec.~\ref{Sec:4}, 
we conclude the paper and discuss possible applications and perspectives. 
In appendices, we provide more detailed accounts of the model and analytic computations.

\sect{Formulation} \label{Sec:2}
Nonperturbative phenomena, such as the formation of condensates, can be studied by the mean-field method (See Refs.~\cite{Read:1983,Coleman:1983}  for early works on the Kondo effect).
In Refs.~\cite{Yasui:2016svc,Yasui:2017izi},  
the QCD Kondo effect was investigated with 
the mean-field approach applied to 
a four-point interaction model 
analogous to the Nambu--Jona-Lasinio (NJL) model (for applications, see Refs.~\cite{Suzuki:2017gde,Yasui:2017bey,Fariello:2019ovo,Suenaga:2019car,Suenaga:2019jqu,Kanazawa:2020xje,Araki:2020fox,Suenaga:2020oeu,Ishikawa:2021bey,Suenaga:2021wio}).

Referring to those studies, 
we extract the essence of the chiral symmetry breaking and the Kondo effect in the Dirac fermion systems in strong magnetic fields.  
For this purpose, we use the following Lagrangian in the $(3+1)$-dimensional spacetime 
(See Appendix~\ref{sec:model}) 
\begin{align}
{\cal L} =& \bar{\psi} (i \sla \pd_\para - m_l) \psi + \frac{G_{ll}}{2N}  \left[ (\bar{\psi} \psi)^2 + (\bar{\psi} i\gamma_5 \psi)^2  \right] 
\label{eq:L2}
\\
& +  \sum_{c = \pm}  \Big[ \,  c \bar{\Psi}_v^c i \partial_0 \Psi_v^c
\nnb
& + \frac{G_{hl}}{N} \left\{ ( \bar{\psi} \Psi_v^c ) (\bar{\Psi}_v^c \psi) 
+ ( \bar{\psi} i\gamma_5 \Psi_v^c )( \bar{\Psi}_v^c i\gamma_5 \psi )  \right\}  \, \Big]
\nn
,
\end{align}
where $\gam_5 \equiv i \gam^0 \gam^1 \gam^2 \gam^3$. 
This Lagrangian contains four eigenmodes of $  \psi  $ for 
the spin-polarized light fermion and its antifermion in the lowest Landau level (LLL), where $ m_l $ is the light-fermion mass. 
It has the index for the inner degrees of freedom such as the color for quarks ($N=3$) and the pseudospin in condensed matter, e.g., graphene ($N=2$). 
We assume that a magnetic field is applied in the third ($z$) direction. 
Then, $  \psi  $ has the kinetic term $ \sla \pd_\para \equiv \gam^0 \pd_0 + \gam^3 \pd_3  $  and is an eigenstate of 
the spin along the magnetic field due to the Zeeman effect, 
i.e., $ i \gam^1 \gam^2 \psi = \sgn(q_l B) \psi $ with an electric charge $  q_l$. 
Note that planar systems such as graphene can be discussed with straightforward modifications of the present formulation.

The Lagrangian \eqref{eq:L2} also contains 
$ \Psi_v^+ $ and $ \Psi_v^- $ for 
the heavy-fermion and its antifermion fields introduced as impurities.  
They stem from the original Dirac field $\Psi$ as $\Psi_v^\pm \equiv e^{ - i \frac{ \sla \pd_\perp}{2m_h}}  e^{ \pm i m_h v^\mu x_\mu } \Q_\pm  \Psi $ with $\sla \pd_\perp \equiv  (\gam^\mu - \sla v v^\mu) \pd_\mu$ after the factorization of the plane-wave components at the position $x^\mu$ and the projection by $ \Q_\pm = \frac{1}{2} (1 \pm v^\mu \gamma_\mu)$ as explained in \suppA.  
Here, the four-momentum is given as $ m_h v^\mu $ with $m_h$ and $v^\mu$ being the heavy-fermion mass and the four velocity, respectively. 
In the rest flame of heavy fermions, $v^\mu=(1,0,0,0)$, $\Q_\pm = \frac{1}{2} (1 \pm \gamma_0)$ is the projection operator to the particle and antiparticle components.\footnote{
This treatment is similar to but slightly different from the conventional framework of the heavy-quark effective theory (HQET)~\cite{Eichten:1989zv,Georgi:1990um,Neubert:1993mb,Manohar:2000dt} in high-enerygy physics, which is an effective field theory for heavy quarks in QCD.
In the conventional HQET, one retains either $ \Psi_v^+ $ or  $ \Psi_v^- $. 
On the other hand, in this work we retain both of $ \Psi_v^+ $ and $ \Psi_v^- $ to include the heavy fermion and antifermion in a charge-conjugation invariant manner~\cite{Korner:1991kf,Balk:1993ev}. 
\shorten{
This can be derived by the Foldy-Wouthuysen transformation~\cite{Korner:1991kf,Balk:1993ev}. 
}
} 
$ G_{hl} >0$ and $ G_{ll} >0$ are the four-point coupling constants for the interactions between the heavy and light fermions
and between the light fermions, respectively. 

\cout{
Notice that the Lagrangian (\ref{eq:L2}) is invariant under the phase rotation 
$ U(2)_V \times U(2)_A  $ operating on the spinor $  (\psi, \Psi_v^c)  $ 
when $  v^\mu = (1,0,0,0)$, $  m_l = 0$, and $ G_{ll} = G_{hl} $. 
\com{This is not true in the absence of the heavy-heavy interactions. 
The HQ limit may be taken earlier.} 
This is an extended isospin symmetry emerging in the heavy-quark limit 
that is mixing between the light and heavy quarks. 
We will see that the Kondo effect in magnetic fields enjoys this emergent isospin symmetry, 
in contrast to the conventional Kondo effect at finite chemical potential 
that breaks the symmetry. 

}

We analyze the Lagrangian (\ref{eq:L2}) with the mean-field approximation. 
\shorten{
We keep the linear fluctuations around the minimum of the effective potential at the one-loop level. 
}
Analogously to Ref.~\cite{Araki:2020fox} in the (3+1)-dimensional spacetime, the chiral and Kondo condensates are assumed to be
\begin{align}
\langle \bar{\psi} \psi \rangle_\mathrm{LLL} \equiv - \frac{N}{G_{ll}} M, 
\quad
\langle \bar{\psi} \Psi_v^\pm \rangle_\mathrm{LLL} \equiv \frac{N}{G_{hl}} \Delta
 \label{MFA_LLL} 
 ,
\end{align}
with the gaps $M$ and $\Delta$.
By explicitly diagonalizing the mean-field Lagrangian, 
we obtain the energy-momentum dispersion relations of the four eigenmodes  
\begin{subequations}
\begin{align}
E_\pm(p_{z}) &\equiv \pm \frac{1}{2} \left(  \sqrt{ E_{p_{z}}^2 +  |2 \Delta|^2 } \pm E_{p_{z}}  \right)
, \\
\tilde{E}_\pm(p_{z}) &\equiv \pm \frac{1}{2} \left( \sqrt{ E_{p_{z}}^2 +  |2 \Delta|^2 } \mp E_{p_{z}} \right),
\end{align}
\end{subequations}
where $E_{p_{z}}\equiv \sqrt{p_z^2+(m_l+M)^2}$. 
Straightforwardly, we obtain the thermodynamic potential $ \Omega (M,\Delta)
\equiv - T \ln Z$ with temperature $T=1/\beta$ 
and the partition function $Z$ at the one-loop level \cite{Kapusta:2006pm}: 
\cout{
\begin{align}
\tilde \Omega (M,\Delta) 
&= \frac{1}{2 \tilde G_{ll}} M^2 + \frac{1}{ 2 \tilde G_{hl}}  |2 \Delta|^2 
\label{Omega}
\\
& -  \sum_{i }
 \int_{-\Lambda}^\Lambda \frac{dp_z}{2\pi} 
 \left[\frac{1}{2} |E_i |
 + \frac{1}{\beta} \ln    (1+e^{-\beta |{E}_i|}) \right]
 \nn
 ,
\end{align}
} 
\cout{
\begin{align}
&
\tilde \Omega (M,\Delta) 
= \frac{1}{2 \tilde G_{ll}} M^2 + \frac{1}{ 2 \tilde G_{hl}}  |2 \Delta|^2 + \tilde \Omega_\vac +  \tilde \Omega_T ,
\nn
\\
&&\tilde \Omega_\vac(M,\Delta) = -  \sum_{{\cal E}_i }
 \int_{-\Lambda}^\Lambda \frac{dp_z}{2\pi}  \frac{1}{2} |{\cal E}_i | ,
 \label{Omega}
 \\
 &&\tilde \Omega_T(M,\Delta) 
 =
  -  \sum_{{\cal E}_i }
 \int_{-\infty}^\infty \frac{dp_z}{2\pi} 
 \frac{1}{\beta} \ln    (1+e^{-\beta |{\cal E}_i|}) ,
 \nn
\end{align} 
} 
\begin{align}
\Omega (M,\Delta) 
&=  N \left[ \, \frac{ M^2}{2 G_{ll}}  
+ \frac{  |2 \Delta|^2}{ 2 G_{hl}}  
+ \Omega_\vac +   \Omega_T  \, \right] ,
\nn
\\
  \Omega_\vac(M,\Delta) &= - \rho_B \sum_{{\cal E}_i }
 \int_{-\Lambda}^\Lambda \frac{dp_z}{2\pi}  \frac{1}{2} |{\cal E}_i | ,
 \label{Omega}
 \\
   \Omega_T(M,\Delta) 
 &=
  - \rho_B \sum_{{\cal E}_i }
 \int_{-\infty}^\infty \frac{dp_z}{2\pi} 
 \frac{1}{\beta} \ln    (1+e^{-\beta |{\cal E}_i|}) ,
 \nn
\end{align}  
where $ \Omega_{\vac/T}$ are  
the contributions from the vacuum bubbles 
and the thermal excitation, respectively, 
and $ \rho_B= |q_l B|/(2\pi) $ is the Landau degeneracy factor. 
The sum runs over the four eigenmodes $ {\cal E}_i = \{ E_+ , E_-,  \tilde{E}_+ , \tilde{E}_-\} $.  
\cout{
Here, we have introduced dimensionless quantities 
$ \tilde \Omega = \Omega/(N \rho_B) $, 
$\tilde G_{hl} =  \rho_B G_{hl}$, and $\tilde G_{ll} =  \rho_B G_{ll} $ with the Landau degeneracy factor $ \rho_B= |q_l B|/(2\pi) $. 
}
We are left with evaluating the one-dimensional momentum integral whose interval is limited within the ultraviolet cutoff $\Lambda$.  
All the dimensionful quantities can be scaled by $\Lambda$ 
because of the dimensional reduction.

\cout{As a result, we are left with evaluating the thermodynamic potential (\ref{Omega}) with the one-dimensional momentum integral \cgd{with $\Lambda$ being the ultraviolet cutoff.}}

The thermodynamic potential (\ref{Omega}) is 
inevitably unstable at the origin $M = 0 = \Delta$ 
due to the dimensional reduction at zero temperature, 
which implies that the system always favors nonzero condensates. 
This can be confirmed with an analytic expression of $\Omega_\vac $ as follows. 
We take the massless limit ($m_l =0$) for a demonstration. 
One can simply perform the $p_z$ integral 
and expand the result as 
\begin{align}
\label{eq:log-vac}
 \Omega_\vac (\R) 
= - \frac{\rho_B}{2\pi}
\biggl( 
\Lambda^2 +  \R^2
\biggl(
\frac12 + \ln  \frac{2\Lambda}{\R}
\biggr) 
+ \order( \R^4 )
\biggr)
,
\end{align}
where $  \R^2 = M^2 + |2\Delta|^2 $. 
Stability at the origin is determined by the sign of the quadratic terms in $\R$. 
The quadratic term enhanced by a logarithmic factor $  \ln \Lambda/\R $, diverging as $ \R \to 0 $,
makes the potential convex upward at the origin for {\it any coupling strengths} 
\cite{Gusynin:1994re,Gusynin:1994va,Gusynin:1994xp,Gusynin:1995nb, Fukushima:2012xw}. 
At finite temperature, the origin can be a stable point since the thermal contribution $ \Omega_T $ yields a term that can cancel the above-mentioned logarithmic term. 

\cout{

\com{
Note:
\begin{align}
\tilde \Omega (M,\Delta) 
\=   \frac{N}{2 \tilde G_{ll}}  m_l^2    
\nnb
&&
-  \frac{N}{2 \tilde G_{ll}} 2 m_l (M+m_l) 
\nnb
&&
 + \frac{N}{2 \tilde G_{ll}} (M+m_l)^2 + \frac{N}{ 2 \tilde G_{hl}}  |2 \Delta|^2 
\nnb
&& - N \sum_{i }
 \int_{-\Lambda}^\Lambda \frac{dp_z}{2\pi} \left[\frac{1}{2} E_i 
 + \frac{1}{\beta} \ln    (1+e^{-\beta {E}_i}) \right]
 \nn
 ,
\end{align}
}

}

\sect{Zero-temperature results} \label{Sec:3} 
We first investigate the ground state at zero temperature.
We consider several cases classified by the relative magnitude between $G_{hl}$ and $G_{ll}$ in Eq.~(\ref{eq:L2}). 
When $G_{hl}=G_{ll}$, the thermodynamic potential (\ref{Omega}) is invariant under a rotation in the $M$-$\Delta$ plane in the chiral limit, i.e., the vanishing light-fermion mass limit ($m_l=0$), 
and has a degenerate minimum on a circle at $\Phi^2 = {\rm constant}$. 
A finite $ m_l$ breaks the rotational invariance. 
As a result, the potential is tilted toward the $M$ axis,  
and has a minimum at a nonzero $M$ and vanishing $\Delta$. 
This is an analog of the chiral condensate in the effective models of QCD with the nonzero current quark mass (see, e.g., Ref.~\cite{Hatsuda:1994pi}). 
When $G_{hl} \neq G_{ll}$, 
there is no longer the degeneracy on the circle. 
When $G_{hl} < G_{ll}$, 
the tendency that $M$ is favored over $\Delta$ is enhanced due to the suppression of the potential energy along the $M$ axis for any $m_l$.  
However, when $G_{hl} > G_{ll}$, competition appears between the suppression of the potential energy and the tilting effect. 
This is a nontrivial case as we investigate in detail below.

\begin{figure}[t!]
    \begin{minipage}[t]{1.0\columnwidth}
        \begin{center}
            \includegraphics[clip, width=1.0\columnwidth]{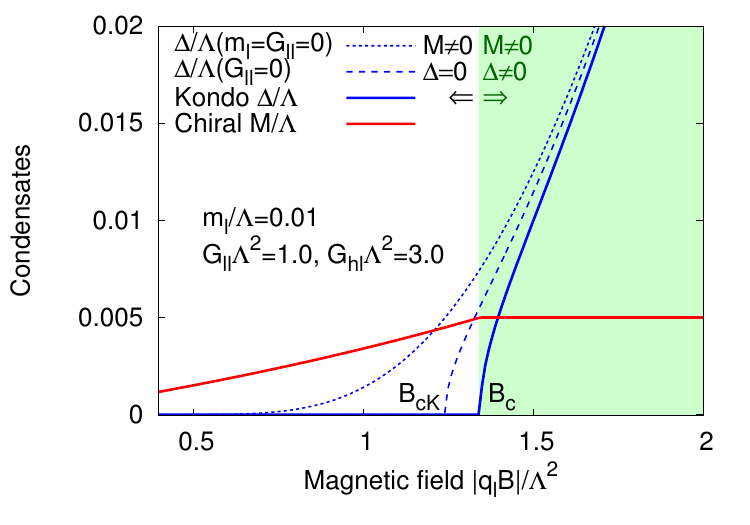} 
        \end{center}
    \end{minipage}
    \vspace{-0.8cm}
    \caption{
Magnetic-field dependences of Kondo condensate $\Delta$ (blue) and chiral condensate $M$ (red).
The solid lines show the full results with fixed parameters, $m_l/\Lambda=0.01$, $G_{ll}\Lambda^2=1.0$, and $G_{hl}\Lambda^2=3.0$.
The dotted and dashed lines show $\Delta$ with $m_l=G_{ll}=0$ and with $G_{ll}=0$, respectively.
 }
\label{fig:B-dep}
\end{figure}

First, we consider a special case $G_{hl} > G_{ll} =0$ 
and elucidate the effects of the light-fermion mass $ m_l $ 
on the Kondo effect.   
In the chiral limit, we find that there is no phase transition characterized by a singular behavior of $\Delta$, as shown by the dotted line in Fig.~\ref{fig:B-dep}. 
\shorten{
Such a monotonic increase is similar to that found at finite chemical potential by the mean-field approach~\cite{Yasui:2016svc}. 
}
On the other hand, when a nonzero $m_l$ is switched on, 
the monotonic increase changes to a second-order phase transition, as shown by the blue dashed line, at 
a critical magnetic-field strength $ B_{cK} $. 
Thus, the light-fermion mass plays an important role in the Kondo effect.

Next, we investigate the competition between the chiral and Kondo condensates with nonzero $ G_{ll} $ and $ G_{hl} $. 
One can find analytic solutions of the stationary conditions 
$\partial  \Omega/\partial M = 0 = \partial   \Omega/\partial \Delta $ as follows 
(see \suppB for details). 
One immediately notices a trivial solution $ \Delta=0$ accompanied by a nontrivial solution $ M \neq 0$. 
This set is denoted as $ (M_1,\Delta_1 = 0) $. 
One can find another set of solutions $ (M_2,\Delta_2)$ as  
\begin{subequations}
\label{eq:sol-2}
\begin{align}
\label{eq:sol-2-chiral}
M_2 &= \frac{m_l G_{ll}}{G_{hl}-G_{ll}} , 
\\
\label{eq:sol-2-Kondo}
\Delta _2 &= 
 \frac{1}{2} \sqrt{  \frac{ \Lambda^2 }{ \sinh^2 \bigl( \frac{2 \pi^2 }{ \rho_B  G_{hl} } \bigr)} 
 - \frac{  m_l^2 G_{hl}^2}{(G_{hl}-G_{ll})^2}   } 
 .
\end{align}
\end{subequations} 
Both the two sets, $ (M_1,\Delta_1 = 0) $ and $ (M_2,\Delta_2)$, satisfy the stationary conditions 
for any values of the parameters as long as $ G_{hl} > G_{ll} $.

\begin{figure}[t!]
    \begin{minipage}[t]{1.0\columnwidth}
        \begin{center}
            \includegraphics[clip, width=1.0\columnwidth] {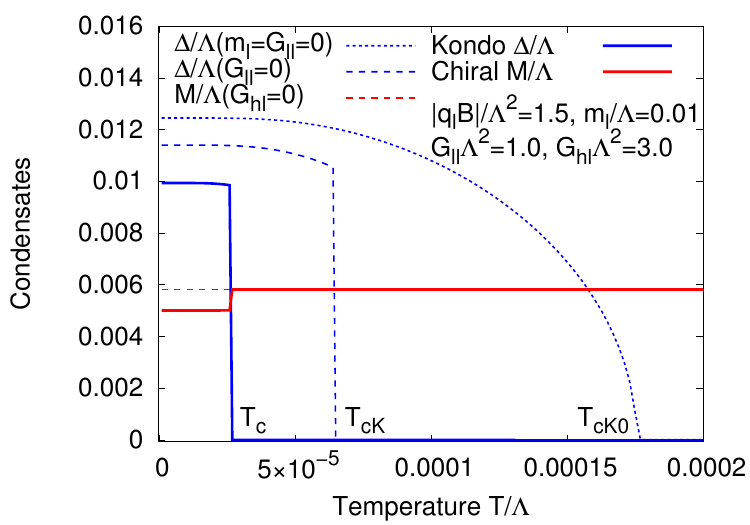}
        \end{center}
    \end{minipage}
    \vspace{-0.8cm}
 \caption{Temperature dependences of Kondo condensate $\Delta$ and chiral condensate $M$ at a fixed magnetic-field strength $|q_lB|/\Lambda^2=1.5$.
The legends are the same as in Fig.~\ref{fig:B-dep}.
The red dashed line shows $ M $ at $ G_{hl} =0$. 
}
\label{fig:T-dep}
\end{figure}
\begin{figure*}[t!]
\includegraphics[clip, 
width=1.8\columnwidth 
]{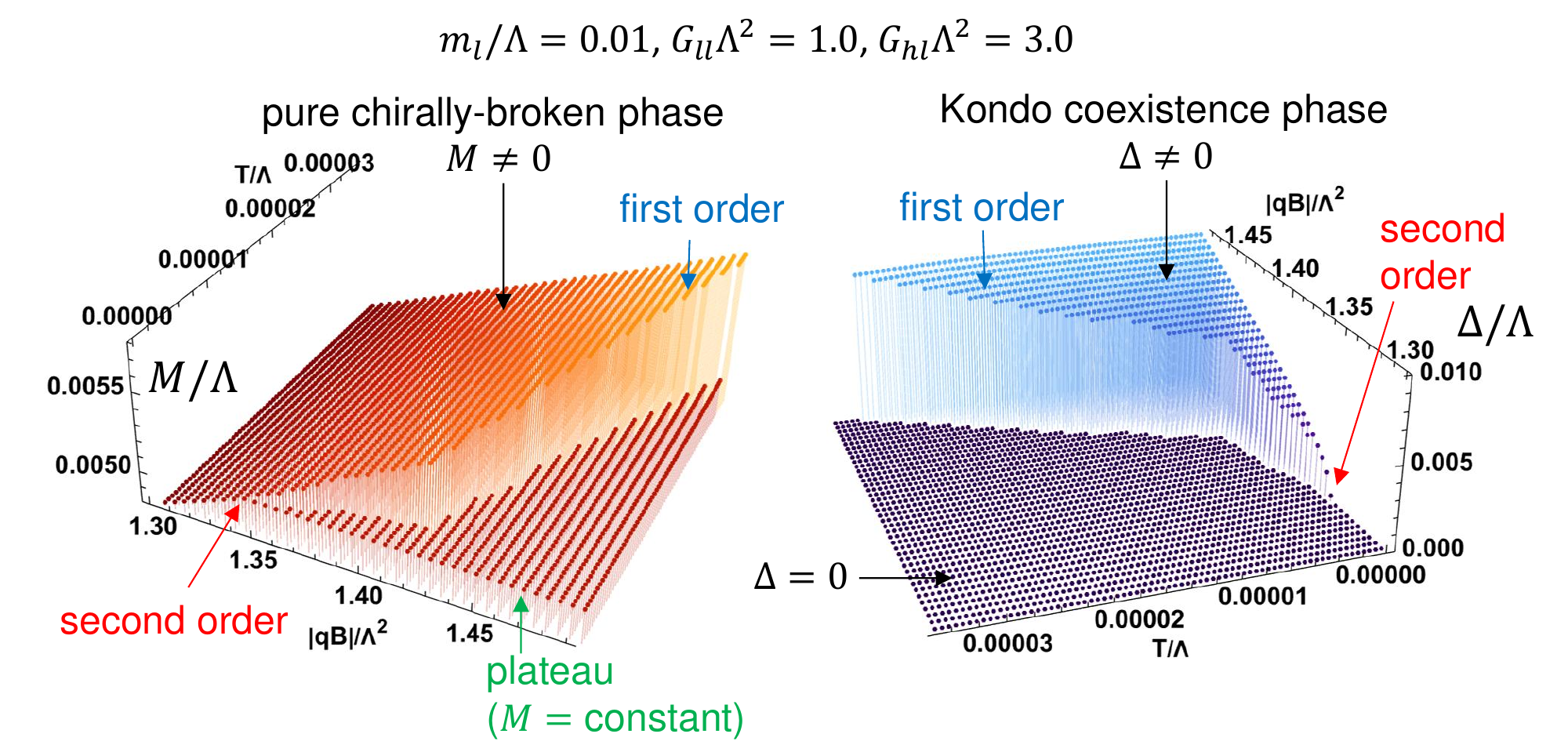}
 \caption{Magnetic-field and temperature dependences of chiral condensate $M$ (left) and Kondo condensate $\Delta$ (right).
}
\label{fig:T-B-dep}
\end{figure*}

We numerically investigated which, or any other set of solutions, serves as 
the global minimum of the thermodynamic potential $  {\Omega}(M,\Delta) $. 
The numerical results are shown with the solid lines in Fig.~\ref{fig:B-dep}.
We find that, below a certain critical magnetic-field strength $ B_c $, the first solutions, $ (M_1,\Delta_1 = 0) $, are realized: The Kondo effect is prohibited by the existence of the chiral condensate.
By solving $\Delta_2=0$ in Eq.~(\ref{eq:sol-2-Kondo}), we obtain the critical strength of magnetic field: 
\begin{align}
q_l B_c = \frac{ 4\pi^3}{ G_{hl} \, \arcsinh \Bigl(   \Lambda  \frac{ |G_{hl}-G_{ll}|}  {  m_l G_{hl}} \Bigr) }
\sim  \frac{ 4\pi^3 m_l }{  \Lambda  |G_{hl}-G_{ll}| }
\label{eq:Bc}
,
\end{align}
where the rightmost side holds with a small value of $|G_{hl}-G_{ll}|$. 
The true vacuum switches over from $ (M_1,\Delta_1 = 0) $ to $ (M_2,\Delta_2)$ at $ B_c $, 
which is thus identified as a quantum critical point.  
The analytic solution~(\ref{eq:sol-2}) shows a nontrivial behavior of the condensates above $B_c$: 
The Kondo condensate grows up as the magnetic field increases, while  the chiral condensate is forced to be a constant value ($M_2$). 
This is an anomalous saturation of the chiral condensate due to the formation of the Kondo condensate. 
As in Fig.~\ref{fig:B-dep}, a ``plateau" of the chiral condensate above $B_c$ serves as a clear signal of the appearance of the Kondo condensate at zero temperature.

\cout{Note: 

\begin{subequations}
\begin{align}
M_2 \= -m_l + \frac{m_l G_{hl}}{G_{hl}-G_{ll}} , 
\\
\Delta _2 \=  \frac12 \sqrt{  \frac{ \Lambda^2 }{ \sinh^2 (2 \pi^2 / \tilde G_{hl} ) } 
 - \frac{  m_l^2 G_{hl}^2}{(G_{hl}-G_{ll}) ^2}   } 
 \nnb
 \=   \frac12 \sqrt{  \frac{ \Lambda^2 }{ \sinh^2 (2 \pi^2 / \tilde G_{hl} ) }   - (M_2 + m_l)^2  } 
 .
\end{align}
\end{subequations}
One can arrange it as 
\begin{align}
(M_2 + m_l )^2 + |2 \Delta_2| ^2 = \frac{ \Lambda^2 }{ \sinh^2 (2 \pi^2 / \tilde G_{hl} ) } 
\end{align}

\begin{align}
1/\rho_B \= \frac{G_{hl}}{2\pi^2} \arcsinh \biggl(   \Lambda  \frac{ |G_{hl}-G_{ll}|}  {  m_l G_{hl}} \biggr)
\nnb
\=   \frac{G_{hl}}{2\pi^2} \arcsinh \biggl(   \frac{  \Lambda } { M_2 + m_l  } \biggr)
\nnb
\nnb
&\sim&  \frac{G_{hl}}{2\pi^2}    \frac{  \Lambda } { M_2 + m_l  }  ,
\end{align}
where the last approximation holds for a small value of $\Lambda$ [a large value of $M_{2}$?].

}

\sect{Finite-temperature results} 
\label{Sec:3-2} 
Finally, we investigate the phase diagram at finite temperature. 
In Fig.~\ref{fig:T-dep}, we show the numerical results for the temperature dependence of the condensates, where the magnetic-field strength and the heavy-light coupling are fixed at $|q_lB|/\Lambda^2=1.5$ and $G_{hl}\Lambda^2 =3.0$, respectively. 
The Kondo condensate at $G_{ll}=0$ and $m_l=0$ is shown by the blue dotted line, where 
we find a phase transition at the critical temperature $T_{cK0}$.
When we switch on a light-fermion mass $ m_l/\Lambda =0.01$, as shown by the blue dashed line, 
we observe that the magnitude of condensate is slightly suppressed, and the critical temperature decreases to $ T_{cK} $. 
This suppression effect by $m_l$ is consistent with our result at zero temperature.

 Before discussing the competition, we check the spontaneous chiral symmetry breaking with nonzero $ G_{ll} $ 
 at $G_{hl}=0 $. 
The chiral condensate $ M $ is shown by the red dashed line in Fig.~\ref{fig:T-dep}.  
The chiral condensate hardly changes 
within the plot range of Fig.~\ref{fig:T-dep} 
but decreases at higher temperature.
The chiral condensate at nonzero $ m_l $ shows a crossover transition, and its pseudocritical temperature is located in the higher-temperature region, not shown in the figure.

The numerical results including both nonzero $ G_{ll}$ and $ G_{hl} $ are shown by solid lines in Fig.~\ref{fig:T-dep}. 
We find that the chiral and Kondo condensates coexist at low temperature below $T_c$. 
Here, the critical temperature $ T_{c} $ for the Kondo condensate decreases from $ T_{cK} $ due to the competition.  
As we increase temperature across $ T_{c} $, 
the Kondo condensate melts away and the chiral condensate starts to {\it increase} abruptly at $ T_{c} $ since the chiral condensate is released from the competition with the Kondo condensate. 
Such a steep increase in the chiral condensate  
can be indirect evidence of the Kondo condensate.  

\begin{figure}[t!]
\includegraphics[clip, width=\columnwidth]{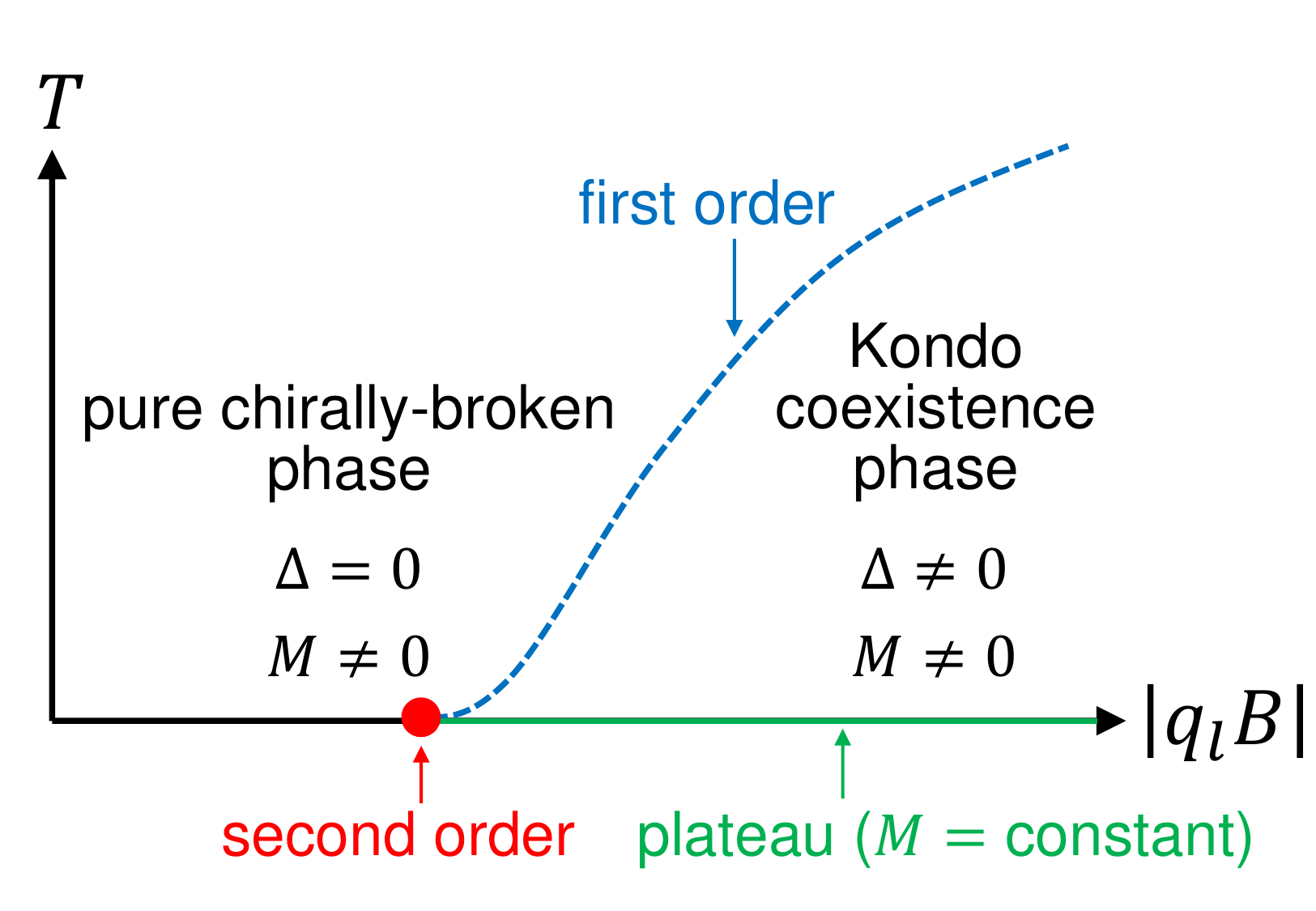}
\vspace{-0.5cm}
\caption{Schematic phase diagram extracted from Fig.~\ref{fig:T-B-dep}.
}
\label{fig:schematic}
\end{figure}

In Fig.~\ref{fig:T-B-dep}, we show the condensate values on the $T$-$B$ plane. 
The plateau, which we have observed in the chiral condensate $M$ at zero temperature, also extends to nonzero temperature. 
In addition, one can read off the orders of phase transitions:
While both condensates indicate second-order phase transitions at zero temperature, 
switching on a nonzero temperature leads to the first-order phase transitions. 
In Fig.~\ref{fig:schematic}, we show a schematic phase diagram, where the first-order phase transition line at finite temperature is connected to the quantum critical point at zero temperature.

\sect{Conclusions} \label{Sec:4} 
We have investigated the novel phase diagram of Dirac fermions, which is characterized by the chiral and Kondo condensates. 
We analytically found the critical magnetic field $B_c$ where the new quantum critical point emerges from the competition between the two condensates. 
This competition also gives rise to a {\it saturation behavior} of 
the chiral condensate above $ B_c$ at low temperature. 
At higher temperature, the Kondo condensate melts away. 
This is the end of the competition, 
which is signaled by  
a steep {\it increase} of the chiral condensate.

For applications, our study can be applied to condensed matter systems as well as quark systems in high-energy physics.
Dirac fermions are known to be realized in a variety of condensed matter systems including graphene. 
The magnetic catalysis in spatial two dimensions has been discussed in Refs.~\cite{Gusynin:1994re,Gusynin:1994va,Gorbar:2002iw, Gusynin:2006gn, herbut2007theory} and was further investigated with lattice simulations for graphene in Refs.~\cite{PhysRevB.89.245404, PhysRevB.95.165442, PhysRevX.6.011029}.  
In addition, a flat band emerges in twisted bilayer graphene with a magic angle due to mixing between the Dirac cones from different layers \cite{bistritzer2011moire, song2022magic, shi2022heavy, chou2023kondo,hu2023kondo} and 
holds the $SU(2)$ valley degeneracy. 
Therefore, stacking this bilayer graphene and another monolayer graphene with a Dirac band will create a platform for the magnetic-field induced Kondo effect.  
Modeling the inter-layer interactions is a hot topic (see, e.g., Ref.~\cite{chen2021electrically}), 
which we leave as a future work.

One of the relevant questions is the excitation spectrum 
when the ground state varies with an increasing magnetic field.  
It is interesting to investigate an extension of the so-called Yu-Shiba-Rusinov state in superconductor \cite{
yu1965bound, shiba1968classical, rusinov1969superconductivity,  rusinov1969theory}
(see also Refs.~\cite{jarrell1990gap, satori1992numerical, *shiba1993numerical, *sakai1993numerical, salkola1997spectral, bauer2013microscopic, kirvsanskas2015yu, lee2017scaling, moca2021kondo} and Ref.~\cite{Kanazawa:2016ihl} for an analog in QCD), 
the Nambu-Goldstone bosons, and 
non-Fermi liquid behavior near quantum critical points \cite{PhysRevB.14.1165, doniach1977kondo, millis1993effect} 
pursued with the experimental progress 
(see, e.g., Refs.~\cite{vojta2003quantum, lohneysen2007fermi, gegenwart2008quantum, si2010heavy}). 
It is important to investigate the critical exponents with possible excitations. 
One can also study an extension to the multi-channel Kondo effect~\cite{Nozieres:1980}, where a non-Fermi liquid system can appear depending on the number of channels (see, e.g., Refs.~\cite{Nozieres:1980, Affleck:1995ge, Kanazawa:2016ihl, Kimura:2016zyv, Kimura:2020uhq}).

We also draw attention to the heavy-quark dynamics in ultrarelativistic heavy-ion collisions \cite{Fukushima:2015wck, Hattori:2016emy}. 
It is an interesting question how the heavy-quark diffusion can be modified by the strong correlation induced by the Kondo effect.

Numerical simulations on the lattice have provided useful approaches to nonperturbative many-body phenomena.  
The intertwined dynamics of the chiral symmetry breaking and the Kondo effect may be simulated in strong magnetic fields. 
Lattice QCD simulations 
\shorten{in strong magnetic fields} 
elucidated the magnetic catalysis in QCD with high precision (see, e.g., Refs.~\cite{Bali:2011qj, Bali:2012zg, Endrodi:2015oba, DElia:2018xwo, Endrodi:2019zrl, Ding:2020hxw, DElia:2021yvk}), and they can be also applied to the counterpart in solid-state physics. 
The magnetically induced Kondo effect deserves further study beyond the mean-field model calculation and should be confirmed by direct measurements of the Kondo condensate constructed from the correlation function of the heavy and light fermions or by indirect ones through special behaviors of the chiral condensate pointed out in this paper. 
Recall also that the systems in a magnetic field and at finite density share analogous low-energy excitations stemming from the degeneracy in the Landau levels and on the Fermi surface. 
Therefore, the Monte Carlo simulation in magnetic fields 
serves as a ``dual'' set-up for that at finite density 
which is highly restricted due to the sign problem.

\begin{acknowledgements}
The authors thank Yasufumi Araki and Kazunori Itakura for helpful discussions.
This work was supported by Japan Society for the Promotion of Science (JSPS) KAKENHI 
under grant Nos. JP17K14277, JP20K14476, 20K03948 and 22H02316, and by the RIKEN special postdoctoral researcher program.
\end{acknowledgements}

\appendix


\section{Model Lagrangian}

\label{sec:model}

Here, we provide a detailed explanation of the model Lagrangian (\ref{eq:L2}) used in the main text. 

\subsection{Light-fermion kinetic term}

The kinetic and mass terms for the light fermions 
is given by the Dirac Lagrangian in the $3+1$ dimensional spacetime 
\begin{eqnarray}
\label{eq:Lag_l}
\Lag_{l} \equiv \bar \psi ( i \sla D - m_l ) \psi
\, ,
\end{eqnarray}
where $\sla D \equiv \gam^\mu D_\mu$ with the Dirac (gamma) matrix $\gam^\mu$ defined by 
the Clifford algebra~\cite{Armitage:2017cjs, fradkin2013field}. 
We assume the Einstein notation for the repeated Lorentz indices $(\mu=0,1,2,3)$. 
The four-component Dirac spinor $\psi$ is composed of the two spin states of particles and antiparticles (holes). 
The covariant derivative, $D^\mu = \pd^\mu + q_l A^\mu_{\rm ext}$, contains the gauge field for an external magnetic field, and satisfies the commutation relation $[i D^1, i D^2] = i q_l B$ with all the other vanishing combinations. Here, without loss of generality, we assume a constant magnetic field in the third direction with the strength $B$. 

We briefly explain how the Lagrangian (\ref{eq:Lag_l}) reduces to that given in Eq.~(\ref{eq:L2}) in the lowest Landau level (LLL). 
By the use of the above commutation relation, 
one can introduce the annihilation and creation operators 
\begin{eqnarray}
a, a^\dagger = \frac{i}{\sqrt{2|q_l B|}}
[D^1 \pm  {\rm sgn}(q_l B) i D^2]
\, ,
\end{eqnarray}
where the upper and lower signs are for $  a$ and $a^\dagger  $, respectively. These operators
satisfy $ [ a,  a^\dagger]=1 $.
One can lower and raise the Landau levels with these operators exactly in the same manner as for the harmonic oscillator.

The Dirac operator is identically cast into the form
\begin{eqnarray}
\label{eq:Lag-Landau}
i \sla D -m_l  =  i \sla \partial _\para -m_l - \sqrt{2|q_l B|} \gam^1 (a \prj_+ +a^\dagger \prj_-)
\, ,
\end{eqnarray}
with the spin-projection operator 
\begin{eqnarray}
    \prj_\pm =  \frac12 (1 \pm i \sgn(q_l B)  \gam^1 \gam^2 ) \, .
\end{eqnarray}
In Eq.~(\ref{eq:Lag-Landau}), the subscript of $ \sla \pd_\para$ means the parallel direction along the magnetic field, i.e., $ \sla \pd_\para \equiv \gam^0 \pd_0 + \gam^3 \pd_3 $. 
The LLL is the ground state that is annihilated by the operator $a$ and is a spin eigenstate with the lower Zeeman energy projected by $\prj_+$. 
Therefore, the last term in Eq.~(\ref{eq:Lag-Landau}) vanishes for the LLL, and we are left with the Lagrangian (\ref{eq:L2}).

It is instructive to see the Landau levels explicitly. 
The Klein-Gordon operator reads 
\begin{eqnarray}
&& \hspace{-1cm}
(i \sla D + m_l )(i \sla D -m_l ) \psi_\pm
\\
\=
- \left[\, \partial_t^2 - \partial_z^2 + (2  a^\dagger a  + 1 \mp 1)  \vert q_l B \vert  + m_l^2 \, \right] \psi_\pm 
\nn
\, ,
\end{eqnarray}
where we defined the spin eigenstates $ \psi_\pm=\prj_\pm\psi $. 
We readily find the energy levels $E^2 = p_z^2 + (2n + 1 \pm 1)\vert q_l B \vert$ with a non-negative integer $n \geq 0$ when the Klein-Gordon equation is satisfied by $\psi_\pm$. 
Note that the dispersion relation for the LLL, and thus the kinetic term in the Lagrangian (\ref{eq:L2}), do not depend on the magnetic-field strength $B$.

\subsection{Heavy-fermion kinetic term}

We also briefly summarize the derivation of 
the low-energy heavy-fermion kinetic term in terms of the large-mass expansion \cite{Manohar:2000dt,Korner:1991kf,Balk:1993ev}. 
We decompose the heavy-fermion momentum as 
\begin{eqnarray}
p^\mu = m_h v^\mu + k^\mu
\label{eq:p_Q}
\, ,
\end{eqnarray}
where the four velocity, 
$v^\mu=(v^{0},{\bm v})$, is normalized as $ v^\mu v_\mu = 1$, and $k^\mu$ is the residual momentum with a scale much smaller than the heavy-fermion mass $m_h$.
We introduce projection operators 
\begin{eqnarray}
    \Q_\pm = \frac12 (1 \pm \sla v )
    \, ,
\end{eqnarray}
with $\sla v = \gam^\mu v_\mu $. 
In the rest frame ($ {\bm v}=0 $), these operators project out the particle and antiparticle states. 
The low-energy excitation will be an eigenstate of $\Q_\pm$ since the coupling between the particle and antiparticle states is highly suppressed by the large mass.

\cout{
Accordingly, we project out the eigenstates of $\Q_\pm $ and factorize the plane-wave component as 
\begin{eqnarray}
\Psi_v^\pm (x) \equiv 
e^{ \pm i m_h v^\mu x_\mu} \Q_\pm \Psi(x)
\, ,
\end{eqnarray}
where $\Psi(x)$ is the four-component Dirac spinor of the heavy fermion. 
Then, we expand the heavy-fermion field as 
\begin{eqnarray}
\Psi =  e^{- i m_h v^\mu x_\mu} 
\pp{e^{ i \frac{ \sla \pd_\perp}{2m_h}} } \Psi_v^+  
+  e^{+ i m_h v^\mu x_\mu} 
\pp{e^{ i \frac{ \sla \pd_\perp}{2m_h}} } \Psi_v^- 
\, ,
\end{eqnarray}
}

One can organize a low-energy effective theory in terms of the $1/m_h$ expansion in such a way that there is no mixing between the eigenstates of $\Q_\pm$~\cite{Korner:1991kf,Balk:1993ev}. 
To get the leading-order term in the expansion, one can factorize the heavy-fermion field $\Psi$ as 
\begin{eqnarray}
\Psi \=  e^{- i m_h v^\mu x_\mu} 
e^{ i \frac{ \sla \pd_\perp}{2m_h}}  \Psi_v^+  
+  e^{+ i m_h v^\mu x_\mu} 
e^{ i \frac{ \sla \pd_\perp}{2m_h}}  \Psi_v^- 
\nnb
\= e^{- i m_h v^\mu x_\mu} 
( 1 + \frac{ i\sla \pd_\perp}{2m_h} ) \Psi_v^+  
\nnb&&
+  e^{+ i m_h v^\mu x_\mu} 
( 1 +  \frac{ i \sla \pd_\perp}{2m_h} ) \Psi_v^- 
+ \order(m_h^{-2})
\label{eq:HQ-expansion}
\, .
\end{eqnarray} 
In other words, one can define eigenstates of $\Q_\pm$ as $\Psi_v^\pm \equiv e^{ - i \frac{ \sla \pd_\perp}{2m_h}}  e^{ \pm i m_h v^\mu x_\mu } \Q_\pm  \Psi $ that satisfy $ \Q_\pm \Psi_v^\pm = \Psi_v^\pm$. 
The exponential factor works for the leading-order result with $\sla \pd_\perp \equiv  (\gam^\mu - \sla v v^\mu) \pd_\mu$, and needs to be improved on an order-by-order basis (see Refs. \cite{Korner:1991kf,Balk:1993ev}).  
Multiplying the above expansion (\ref{eq:HQ-expansion}) by the Dirac operator, we obtain an expansion of the heavy-fermion kinetic term 
\begin{eqnarray}
\Lag_h &\equiv& \bar \Psi  (i\sla \pd -m_h) \Psi 
\nnb
\= \sum_{c=\pm}
c \bar \Psi_v^c  \left( \, i v^\mu \pd_\mu  \, \right) \Psi_v^c + \order (m_h^{-1})
\label{eq:Lh}
\, ,
\end{eqnarray} 
where we used identities 
\begin{eqnarray}
\label{eq:Qid} 
&&
\Q_\pm \Q_\pm = \Q_\pm
\, , \quad 
\Q_\pm \Q_\mp = 0
\, ,  \quad
\Q_\pm \gam^\mu \Q_\pm = \pm v^\mu \Q_\pm
\, , \quad
\nnb
&&
\Q_\pm \gam^\mu \Q_\mp 
= (  \gam^\mu  -   \sla v v^\mu ) \Q_\mp 
 .
\end{eqnarray}
The rightmost side in Eq.~(\ref{eq:Lh}) 
is used in the Lagrangian (\ref{eq:L2}) as the low-energy heavy-fermion kinetic term.

\cout{

\com{[NOTE]

\begin{eqnarray}
\Lag_h \= \bar \Psi 
[ e^{- i m_h v^\mu x_\mu} 
(m_h \sla v + i\sla \pd -m_h) \Q_+ \Psi
\nnb &&
+  e^{+ i m_h v^\mu x_\mu}  
(- m_h \sla v + i\sla \pd -m_h) \Q_- \Psi ]
\nnb 
 \= \bar \Psi 
[ e^{- i m_h v^\mu x_\mu} 
( i\sla \pd - 2 m_h \Q_-) \Q_+ \Psi
\nnb &&
+  e^{+ i m_h v^\mu x_\mu}  
( i\sla \pd - 2 m_h Q_+) \Q_- \Psi ]
\nnb 
 \= 
\bar \Psi \Q_+   ( i\sla \pd) \Q_+ \Psi 
+ \bar \Psi \Q_-  ( i\sla \pd) \Q_- \Psi 
\nnb 
&& +  
[ e^{- 2 i m_h v^\mu x_\mu} \bar \Psi \Q_- ( i\sla \pd) \Q_+ \Psi 
+  e^{+ 2i m_h v^\mu x_\mu} \bar \Psi \Q_+ ( i\sla \pd) \Q_- \Psi ]
\nnb 
 \= 
\bar \Psi \Q_+   ( i v^\mu \pd_\mu ) \Q_+ \Psi 
- \bar \Psi \Q_-   ( i v^\mu \pd_\mu ) \Q_- \Psi 
\\
&& +  
[ e^{- 2 i m_h v^\mu x_\mu} \bar \Psi 
i ( \gam^\mu  -  v^\mu \sla v )  \pd_\mu \Q_+ \Psi 
\nnb
&& +  e^{+ 2i m_h v^\mu x_\mu} \bar \Psi 
i (\gam^\mu  -  v^\mu \sla v ) \pd_\mu \Q_- \Psi ]
\nn
\end{eqnarray}

The LO from the NLO-operator term reads 
\begin{eqnarray}
\Lag_h^{\rm NLO} \= \bar \Psi 
[ e^{- i m_h v^\mu x_\mu} 
(m_h \sla v + i\sla \pd -m_h) 
\cgd{\frac{i \sla D_\perp}{2m_h}} \Q_+ \Psi
\nnb &&
+  e^{+ i m_h v^\mu x_\mu}  
(- m_h \sla v + i\sla \pd -m_h) 
\cgd{\frac{i \sla D_\perp}{2m_h}}  \Q_- \Psi]
\nnb
\= \bar \Psi 
[ e^{- i m_h v^\mu x_\mu} 
(-\Q_-) \cgd{ i\sla D_\perp} \Q_+ \Psi
\nnb &&
+  e^{+ i m_h v^\mu x_\mu}  
(- \Q_+) \cgd{i\sla D_\perp}  \Q_- \Psi] 
+ \order(\frac{1}{m_h})
\end{eqnarray}
These terms cancel the off-diagonal components in the LO. 

}

\gr{
\begin{align}
\Psi =& e^{- imv \cdot x} \left( e^{\frac{\mathcal{O}_1}{2m}} e^{\frac{\mathcal{O}_2}{2m^2}} e^{\frac{\mathcal{O}_3}{2m^3}} \cdots \right)  \Psi_v^+ \nonumber\\
&+  e^{imv \cdot x} \left( e^{\frac{\mathcal{O}_1}{2m}} e^{\frac{\mathcal{O}_2}{2m^2}} e^{\frac{\mathcal{O}_3}{2m^3}} \cdots \right) \Psi_v^-.
\end{align}
The $\Psi_v^+$ and $\Psi_v^-$ are defined as
\begin{align}
&\Psi_v^+ \equiv  \left( \cdots e^{-\frac{\mathcal{O}_3}{2m^3}} e^{-\frac{\mathcal{O}_2}{2m^2}} e^{-\frac{\mathcal{O}_1}{2m}} \right) e^{i mv \cdot x} \frac{1}{2}(1+ v^\mu \gamma_\mu) \psi_2, \\
&\Psi_v^- \equiv \left( \cdots e^{-\frac{\mathcal{O}_3}{2m^3}} e^{-\frac{\mathcal{O}_2}{2m^2}} e^{-\frac{\mathcal{O}_1}{2m}} \right) e^{-i m v \cdot x}  \frac{1}{2}(1- v^\mu \gamma_\mu) \psi_2,
\end{align}
where the differential operators $\mathcal{O}_i$ are determined order by order in the $1/m$ expansion of the massive Dirac field (see Refs.~\cite{Korner:1991kf,Balk:1993ev} for the explicit forms).

}

\com{
Here are some basic properties of the projection operators 
\begin{eqnarray}
\label{eq:Qid}
&&\hspace{-0.5cm}
\Q_\pm \Q_\pm 
= \Q_\pm
\, , \quad 
\Q_\pm \Q_\mp 
= 0
\, ,
\\
&&\hspace{-0.5cm}
\Q_\pm \gam^\mu \Q_\pm 
= \pm v^\mu \Q_\pm
\, , \quad
\Q_\pm \gam^\mu \Q_\mp 
\cgd{= (  \gam^\mu  -  v^\mu \sla v ) \Q_\mp}
\nn
 .
\end{eqnarray}

\begin{eqnarray}
\Q_\pm \gam^\mu \Q_\mp 
&=& \frac14 ( 1 \pm \sla v) \gam^\mu (1\mp \sla v)
\nnb
\= \frac14 (  \gam^\mu  \pm \sla v \gam^\mu
\mp  \gam^\mu \sla v - \sla v \gam^\mu \sla v)
\nnb
\= \frac14 (  \gam^\mu  \pm \sla v \gam^\mu
\mp  \gam^\mu \sla v 
- (2 v^\mu \sla v- \gam^\mu) )
\nnb
\= \frac12 (  \gam^\mu  -  v^\mu \sla v )
\pm   \frac14 ( \sla v \gam^\mu - \gam^\mu \sla v )
\nnb
\= \frac12 (  \gam^\mu  -  v^\mu \sla v )
\mp   \frac12 ( \gam^\mu  - v^\mu \sla v )  \sla v 
\nnb
\=   (  \gam^\mu  -  v^\mu \sla v ) \Q_\mp
\end{eqnarray}

To get these identities, we used
\begin{eqnarray}
\sla v \sla v 
\= v_Q^\mu v_Q^\nu \{ \gam_\mu, \gam_\mu \}/2 
= 1
\, , 
\\
\sla v \gam^\mu \sla v
\= v^\alpha v^\beta \left( 
g^{\alpha \mu} \gam^\beta + g^{\mu\beta} \gam^\alpha - g^{\alpha\beta} \gam^\mu 
- i \ep^{\lambda \alpha \mu \beta} \gam_\lambda \gam^5
\right)
\nnb
\= 2 v^\mu \sla v- \gam^\mu
\\
\sla v \gam^\mu -  \gam^\mu \sla v 
\= -  2 ( \gam^\mu \sla v - v^\mu) \sla v \sla v 
\nnb
\= -  2 ( \gam^\mu  - v^\mu \sla v )  \sla v 
\, .
\end{eqnarray} 

}

}

\subsection{Interaction Lagrangian} 
\label{sec:mixing_term}

We examine a sequence of the terms generated from an interaction Lagrangian that has the Lorentz and $\text{SU}(N)$ symmetries~\cite{Ebert:1994tv,Yasui:2013xr,Yasui:2016svc,Yasui:2017izi}: 
\begin{eqnarray} 
{\cal L}_{ \text{int} }
=
-G
(\bar{\psi}\gamma^{\mu}T^{a}\psi)(\bar{\Psi}\gamma_{\mu}T^{a}\Psi), 
\label{current_Lagrangian}
\end{eqnarray}
where $\gamma^{\mu}$ ($\mu=0,1,2,3$) are the Dirac matrices, 
and $T^{a}$ ($a=1,2,\dots,N^{2}-1$) are
the generator matrices of the $\text{SU}(N)$ group. 
The light and heavy fermions, $\psi$ and $\Psi$, belong to the fundamental representation of the $\text{SU}(N)$ symmetry. 
The sums over $\mu$ and $a$ are implicitly taken.  
Here, $G>0$ is the coupling constant.
This is a simple analog of the NJL interaction~\cite{Nambu:1961tp, Nambu:1961fr, Klevansky:1992qe, Hatsuda:1994pi}, 
where the inter-fermion interactions 
are understood as the mimic of the gluon exchange interactions between quarks in QCD.

The matrix structures in Eq.~(\ref{current_Lagrangian}) can be decomposed by utilizing the Fierz transformations, i.e.,
\begin{eqnarray}
 (\gamma^{\mu})_{\alpha \beta} (\gamma_{\mu})_{\gamma \delta}
&=&
\delta_{\alpha \delta} \delta_{\gamma \beta}
+ (i\gamma_{5})_{\alpha \delta}  (i\gamma_{5})_{\gamma \beta}
\label{Fierz_Dirac}
\\
&& -\frac{1}{2} (\gamma^{\mu})_{\alpha \delta}  (\gamma_{\mu})_{\gamma \beta}
-\frac{1}{2} (\gamma^{\mu}\gamma_{5})_{\alpha \delta}  (\gamma_{\mu} \gamma_{5})_{\gamma \beta},
\nonumber 
\end{eqnarray}
for the Dirac matrices and
\begin{eqnarray}
\hspace{-0.5cm}
 (T^{a})_{ij}  (T^{a})_{kl}
=
\frac{N^{2}-1}{2N^{2}} \delta_{il}  \delta_{kj}
- \frac{1}{N} (T^{a})_{il}  (T^{a})_{kj},
\label{Fierz_SUN}
\end{eqnarray}
for the $\text{SU}(N)$ generators. 
As we briefly discuss below (cf.~Refs.~\cite{Yasui:2016svc,Yasui:2016yet,Yasui:2017izi}), 
we identify the dominant terms, that are maintained 
in Eq.~(\ref{eq:L2}) in the main text, as the singlet channel $\delta_{il}  \delta_{kj}$ in the $\text{SU}(N)$ generators. 
Then, we have a relation between the coupling strengths, $G_{hl}$ in Eq.~(\ref{eq:L2}) and $G$ in Eq.~(\ref{current_Lagrangian}), as $G_{hl}/N=G(N^{2}-1)/(2N^{2})$. 
As for the Dirac matrices, we focus on 
the scalar $\delta_{\alpha \delta} \delta_{\gamma \beta}$ and pseudoscalar $(i\gamma_{5})_{\alpha \delta}  (i\gamma_{5})_{\gamma \beta}$ terms, 
and potential realization of the vector and axial-vector condensates in magnetic fields are not considered in Eq.~(\ref{eq:L2}). 

We note that the above singlet channel is the most dominant one in the large $N$ limit \cite{Yasui:2016svc,Yasui:2016yet,Yasui:2017izi}, where the mean-field approximation can be justified.\footnote{See, e.g., Refs.~~\cite{Read:1983,Coleman:1983} for early works on the large $N$ limit applied to the Kondo effect and Ref.~\cite{coleman_2015} for generalities.} 
The other channel, that is, the adjoint representation
from the second term on the right-hand side of Eq.~(\ref{Fierz_SUN}), 
is suppressed in comparison to the singlet channel in the large $N$ limit.

\cout{
By applying another Fierz transformation to Eq.~(\ref{current_Lagrangian}) (cf.~ Ref.~\cite{Buballa:2003qv}),
one can get the ``particle-particle" channels ($\psi^{t}\Gamma\Psi$ and $\Psi^{t}\Gamma^{t}\psi$ types) \cgd{as pairings between the light and heavy fermions}, in addition to the ``particle-antiparticle" channel ($\psi\Gamma' \bar{\Psi}$ and $\bar{\Psi}\bar{\Gamma}'\psi$ types) \cgd{in Eq.~(\ref{eq:L2})}. 
}

In addition to the ``particle-antiparticle" channel ($\psi\Gamma' \bar{\Psi}$ and $\bar{\Psi}\bar{\Gamma}'\psi$ types) in Eq.~(\ref{eq:L2}),  
one can get the ``particle-particle" channels ($\psi^{t}\Gamma\Psi$ and $\Psi^{t}\Gamma^{t}\psi$ types) from 
Eq.~(\ref{current_Lagrangian}) as interactions between the light and heavy fermions (see, e.g.,  Ref.~\cite{Buballa:2003qv} for similar discussions in case of light-light fermion interactions). 
Here $\Gamma$, $\Gamma^{t}$ (the transpose of $\Gamma$), $\Gamma'$, and $\bar{\Gamma}'=\gamma_{0}\Gamma'^{\dag}\gamma_{0}$ (the complex conjugate of $\Gamma'$)
are appropriate combinations of the Dirac matrices 
resulting from the Fierz transformations. 
These terms are also suppressed in the large $N$ limit.

\section{Solutions for gap equations} 
\label{sec:gap-eqs}

From Eq.~(4) in the main text, the stationary conditions 
$\partial \Omega/\partial M = 0 = \partial \Omega/\partial \Delta $ lead to the gap equations 
\begin{eqnarray}
\label{eq:gap-eqs}
\frac{M}{G_{ll} } +  \frac{M+m_l}{2\pi} \rho_{B} \ln \Xi = 0 ,   \, \,
\frac{\Delta }{G_{hl}} +  \frac{\Delta }{2\pi} \rho_{B} \ln \Xi  = 0 
,
\end{eqnarray}
where
\begin{eqnarray}
\Xi = \frac{ (M+m_l)^2 + 4|\Delta|^2} 
{( \Lambda  + \sqrt{ (M+m_l)^2 + 4|\Delta|^2 + \Lambda ^2} )^2 }.
\end{eqnarray}
By solving these gap equations, 
we can determine $M $ and $ \Delta $. 

\subsection{Without competition}

We briefly investigate special cases 
where $ G_{hl}=0 $ or $G_{ll} =0 $. 
According to the discussion around Eq.~(5), 
we find nonzero solutions for the gap equations 
at any coupling strengths. 
When $ G_{hl}=0 $ in the Lagrangian (\ref{eq:L2}), the gap equation becomes the first form in Eq.~(\ref{eq:gap-eqs}) with $\Delta=0$.
In the chiral limit $m_l=0$, 
we find an analytic solution ($M>0$):
\begin{align}
 M  = \frac{\Lambda}{\sinh \bigl( \frac{\pi}{\rho_B G_{ll}} \bigr)  }
\to 2 \Lambda e^{ - \frac{\pi}{\rho_B G_{ll}} } 
, \label{M_exa}
\end{align}
which approaches the rightmost side for a small value of $G_{ll}$. 
The chiral symmetry is always broken 
under a strong magnetic field and a nonzero coupling constant 
by the mechanism of {\it magnetic catalysis} \cite{Gusynin:1994re,Gusynin:1994va,Gusynin:1994xp,Gusynin:1995nb}.

\cout{
This nonzero solution \cgd{$M>0$} exits at any finite strength of the attractive interaction ($ G_{ll} >0$) and the nonzero magnetic field,
which indicates the occurrence of the {\it magnetic catalysis} of the chiral symmetry breaking 
\cite{Gusynin:1994re,Gusynin:1994va,Gusynin:1994xp,Gusynin:1995nb}:
The chiral symmetry is always broken under a magnetic field and a nonzero coupling constant.
}

On the other hand, 
when $ G_{ll}=0 $ in the Lagrangian (\ref{eq:L2}), the gap equation becomes the second form of Eq.~(\ref{eq:gap-eqs}) with $M=0$.
In the chiral limit, we find an analytic solution ($\Delta>0$):
\begin{align}
\Delta = \frac{\Lambda}{2\sinh \bigl(  \frac{\pi}{\rho_B G_{hl}}  \bigr)}
\to \Lambda e^{ - \frac{\pi}{\rho_B G_{hl}} }  
, \label{delta_exa}
\end{align}
which approaches the rightmost side for a small value of $G_{hl}$. 
This solution clearly indicates the occurrence of the Kondo effect by a strong magnetic field.
The novel magnetically induced Kondo effect was previously suggested also by a renormalization-group (RG) analysis~\cite{Ozaki:2015sya} and further analyzed by a conformal-field theory approach~\cite{Kimura:2018vxj}. 
Our mean-field solution (\ref{delta_exa}) is consistent 
with the emergent scale in the previous RG analysis \cite{Ozaki:2015sya} 
and determines the magnitude of the Kondo condensate for the first time. 
The magnetic-field dependence of $\Delta$ at $G_{ll}\Lambda^2=1.0$ is shown with the blue dotted line in Fig.~1. 

\subsection{With competition}

We also obtain analytic solutions in the presence of 
the competition between the Kondo effect and the magnetic catalysis with nonzero $ G_{hl} $ and $ G_{ll} $. 
We immediately notice that the gap equations (\ref{eq:gap-eqs}) have a trivial solution $ \Delta=0$. 
In this case, $ M $ takes a nontrivial solution which is an extension of Eq.~(\ref{M_exa}) with a nonzero $  m_l$. 
We call these solutions $ (M_1,\Delta_1 = 0) $. 
By eliminating $\ln \Xi $ in Eq.~(\ref{eq:gap-eqs}), 
one can find another set of solutions that 
we call $ (M_2,\Delta_2)$ and show in Eq.~(6).

\bibliography{reference}

\end{document}